\documentclass[journal]{IEEEtran}

\usepackage{amssymb}
\usepackage{hyperref}
\usepackage{wrapfig, framed}

\usepackage[switch]{lineno}

\ifCLASSINFOpdf
  \usepackage[pdftex]{graphicx}
  \graphicspath{{..}{figures/}}
\else
\fi

\begin{document}

\title{Measurement of Excess Noise in Thin Film\\ and Metal Foil Resistor Networks}

\author{Nikolai Beev% <-this % stops a space
\thanks{N. Beev is with the High Precision Measurement Section, Electrical Power Converters Group, Accelerator Systems Department at CERN (Geneva, Switzerland).
}}% <-this % stops a space

\maketitle

\begin{abstract}

Low-frequency resistance fluctuations cause excess noise in biased resistors. The magnitude of these fluctuations varies significantly between different resistor types. In this work measurements of excess noise in precision thin film and metal foil resistor networks are presented. The lowest levels were found in metal foil devices, followed by thin film NiCr networks deposited on silicon substrates. Higher excess noise with a significant spread between different types was seen in thin film devices deposited on ceramic substrates.

\end{abstract}

\begin{IEEEkeywords}
excess noise, noise index, resistor network, thin film, metal foil, LHC, HL-LHC
\end{IEEEkeywords}

\IEEEpeerreviewmaketitle

\section{Introduction}
\IEEEPARstart{R}{esistors} exhibit two distinctly different noise mechanisms.
Thermal noise, also known as Johnson or Johnson-Nyquist noise, does not depend on the resistor type or construction but only on its physical temperature. It has a frequency-independent power spectrum density, which is commonly expressed as open-circuit \textit{voltage noise} in a Thévenin equivalent circuit (eq. \ref{eq1}) or short-circuit \textit{current noise} in a Norton equivalent circuit (eq. \ref{eq2}). A third special case exists: the power-matching condition \textit{R\textsubscript{source} = R\textsubscript{load}}, where half of the noise power is delivered from the noise-generating source resistor to the load (eq. \ref{eq3}). The third case is presented to emphasize the fact that the thermal noise \textit{power spectrum density} is independent of resistance, unlike the voltage or current spectral densities.

\begin{equation} \label{eq1}
    v_n = \sqrt{4 k_B T R} \quad \left[\frac{V}{\sqrt{Hz}}\ \right]
\end{equation}

\begin{equation} \label{eq2}
    i_n = \sqrt{\frac{4 k_B T}{R}} \quad \left[\frac{A}{\sqrt{Hz}}\ \right]
\end{equation}

\begin{equation} \label{eq3}
    P_n = k_B T \quad
    \left[\frac{W}{Hz}\right]
\end{equation}

Excess noise arises from low-frequency resistance fluctuations and is only observed when a resistor is biased by external voltage or current. It has an \textit{1/f$^\alpha$} power spectrum density, where \textit{$\alpha\approx$1}. Its magnitude depends on many factors related to the resistor technology; it is highest for discontinuous materials such as cermet thick films, and lowest for bulk metallic conductors \cite{zandman} \cite{motchenbacher}. Historically, excess noise is also known as \textit{current noise}, but in this work it is preferred not to use this term due to its ambiguity (see eq. 2).
Excess noise is commonly quantified using the Noise Index (NI) (eq. 4). Although the index is defined in a frequency decade, it is normally expressed simply in dB. For example, a resistor having NI=-40 dB biased with 1 V\textsubscript{DC} would generate 10 nV\textsubscript{rms} in any single decade, 14.1 nV\textsubscript{rms} in two decades ($\sqrt{2}$ higher), 20 nV\textsubscript{RMS} in 4 decades, etc.

\begin{equation} \label{eq4}
    NI = 20 log_{10}\left(\frac{V_{rms}[\mu V]}{V_{DC} [V]}\right) \quad [dB/decade]
\end{equation}

Many different methods for measuring excess noise have been devised \cite{conrad_1960} \cite{seifert} \cite{verbruggen_novel_1989} \cite{moon_digital_1992} \cite{demolder_measuring_1980} \cite{scofield_ac_1987} \cite{lamacchia_measuring_2018}. A technique defined in the 1960s \cite{conrad_1960} was standardized in MIL-STD-202G and IEC 60195, and is still widely used by resistor manufacturers. It is based on rms voltage noise measurement in a passband of 1000 Hz centered at 1000 Hz. The sensitivity of this method is fairly poor, especially for modern low-power precision components having low NI \cite{zandman} \cite{lamacchia_measuring_2018}. As a result, manufacturers typically provide a conservative upper limit for this quantity in the -30 to -40 dB range. One notable exception is the LT5400 resistor network by Analog Devices, which has a specified NI of \textless -55 dB \cite{lt5400}. 

The results presented in this this work were obtained using a sensitive method which is an extension of the technique used by Seifert \cite{seifert}. It enables the measurement of NI below -70 dB for the most commonly used intermediate resistance levels of 1 to 10 k$\Omega$.

Resistor networks offer multiple advantages over discrete components, particularly in circuits where the \textit{ratio of resistances} matters more than their absolute values. The inherently good fabrication matching and thermal coupling of elements within a network lead to tolerance and temperature coefficient (TC) \textit{tracking} specifications that are typically several times better than their absolute counterparts. The simplest integrated networks are three-terminal voltage dividers, and other common configurations include arrays of 4 or 8 independent elements. To the author's knowledge, the present work is the first systematic study of excess noise in such devices.

A skilled circuit designer could gain further advantages from large resistor arrays, as elements from different units can be connected in series or in parallel to achieve one or more of the following design goals: 1) obtain any desired value or ratio; 2) further improve tolerances or TC by statistical averaging; 3) optimize power dissipation on the element or package level; 4) decrease maximum voltage or current per element; 5) reduce the impact of parasitics. Some examples of high precision circuits successfully realized using these methods include 7 V/10 V scaling for buried Zener voltage references \cite{pickering_solid_1995} and current mirrors with sub-ppm TC \cite{bastos_20a_2013} \cite{fernqvist_design_2003}. 

Precision resistor networks based on thin film or metal foil technologies are commercially available from multiple manufacturers. The predominant and most mature thin film technology is based on the Nichrome alloy (NiCr) \cite{rolke_nichrome_1981}. Tantalum Nitride (TaN) resistor networks are also available, the main advantage of this technology being its good tolerance to humidity \cite{bos_performance_1994}. Other less common film types include Chromium Silicon (CrSi), as well as proprietary technologies like \textit{Tetrinox}\textregistered\ (Caddock) and \textit{Tamelox}\textregistered\ (Vishay Dale), the latter being a NiCr/TaN hybrid. The substrate materials used for thin film networks are either Alumina (Al\textsubscript{2}O\textsubscript{3}) or surface-oxidized silicon. The most common packages are plastic surface-mount device (SMD) types, although bare chips and double-substrate "sandwich" structures exist as well. 
Metal foil resistor networks use proprietary NiCr-based alloys. The cold-rolled foil is laminated to a ceramic (Alumina) substrate. Networks are realized either as multiple elements on a single substrate, or as a collection of separate selected chips sharing the same package \cite{prnd1446}. Hermetic metal or ceramic cases exist in addition to plastic SMD packages.

New generations of low-noise voltage references, auto-zero amplifiers, digital-to-analog and analog-to-digital converters facilitate the development of high-performance sources, amplifiers, and DC-coupled measurement systems, which in turn calls for a new critical look at excess noise in precision resistors as a potential contributor of \textit{1/f} noise. Among the motivations for the present work are the development of high-precision digitizers \cite{accnews} and the improvement of the CERN / Metron Designs 10 V/10 mA standards \cite{fernqvist_design_2003} for the needs of the High Luminosity Large Hadron Collider project (HL-LHC) \cite{hilumi}.

\section{Measurement setup}
\subsection{Description}
\begin{figure}[h!] 
\centering
\includegraphics[width=8.5cm]{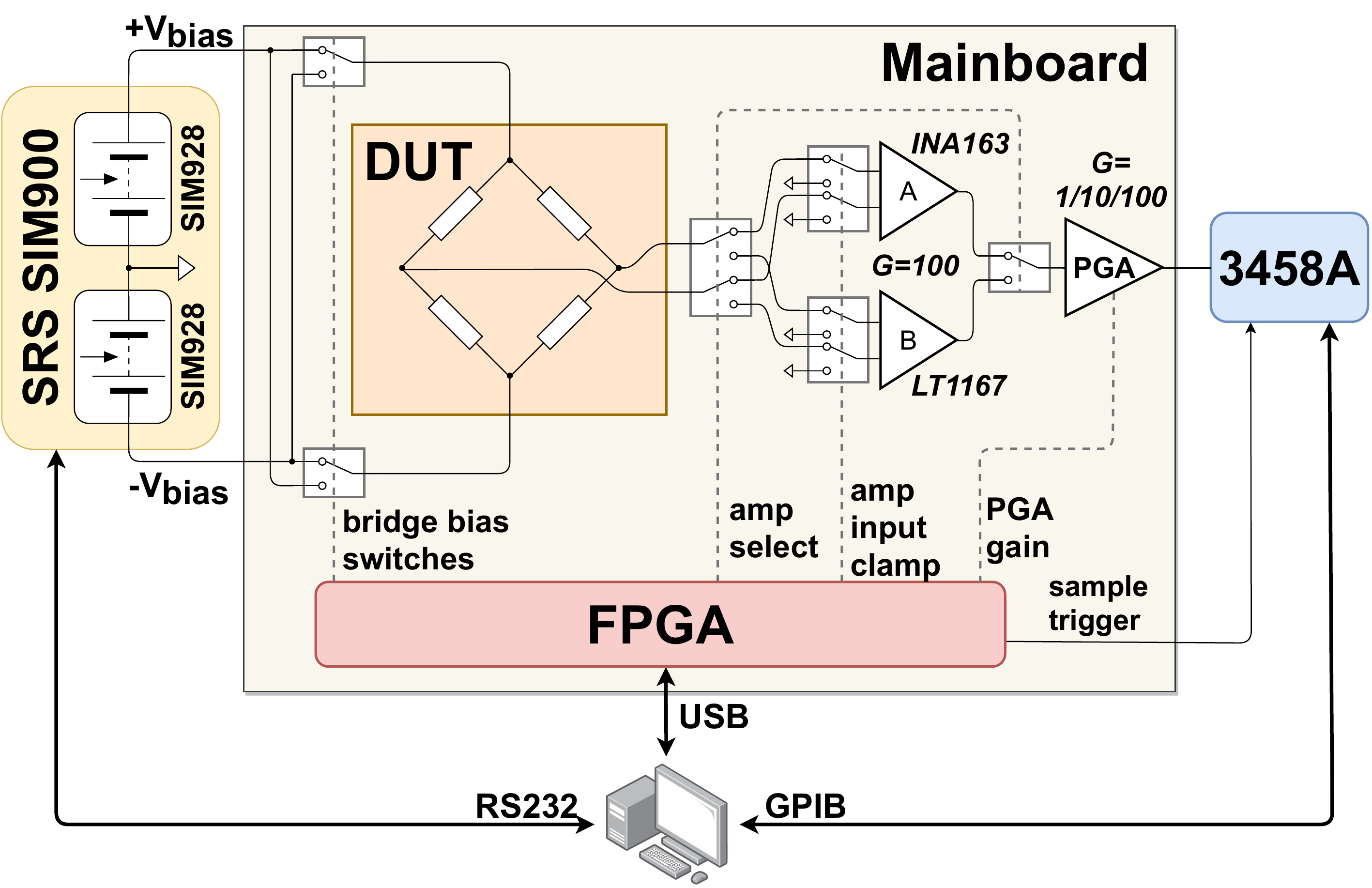}
\caption{Block diagram of the measurement setup}
\label{fig:blockdiagram}
\end{figure}

A block diagram of the measurement setup is shown in Figure \ref{fig:blockdiagram}.

Each Device Under Test (DUT) consisted of one or two resistor networks with their elements connected in a Wheatstone bridge circuit. The DUTs were soldered on connectorized carrier boards that were plugged into the mainboard, which contained the electronics needed for bridge bias voltage reversal, generation of sampling trigger for the digitizer, and amplification of the bridge output signal. All digital signals were generated by a small FPGA module.

Two low-noise battery-based programmable voltage sources \cite{sim928} provided the bias for the bridge. The voltage per element was half of the total Wheatstone bridge bias. It was set to 10 V per element for R\textsubscript{DUT} $\geq$ 2 k$\Omega$ and decreased for lower resistances to stay within the maximum output current of 15 mA. The bias was applied symmetrically, in order to keep the bridge output voltage near zero and thus prevent large common-mode excursions.

The output signal from the bridge was amplified by an instrumentation amplifier. Two such amplifiers were present and were selected by a relay according to the DUT resistance: bipolar-input (INA163 by Texas Instruments) for R\textsubscript{DUT} \textless 10 k$\Omega$ or FET-input (LT1167 by Analog Devices) for R\textsubscript{DUT} $\geq$ 10 k$\Omega$. The inputs of the selected amplifier were clamped for 20 $\mu$s during the bridge bias reversal, in order to prevent saturation from the transient. The output signal of the instrumentation amplifier was further amplified by a programmable gain amplifier (PGA) with gain of 1, 10 or 100.

The amplified signal was digitized by a HP 3458A digital voltmeter (DVM) operated in sampling mode at the 10 V DC range. Sampling was synchronized to the bridge bias reversals and was triggered 20 $\mu$s after the amplifier inputs were taken out of the clamped state. The sampling aperture was set to be as long as possible, in order to prevent the aliasing of broadband noise in the absence of analog Nyquist filtering. For almost all measurements, the maximum total gain of 10000 was used; only in a few cases it was reduced to 1000 to accommodate the output DC voltage to the input range of the DVM. For both gain settings, the noise of the measurement chain exceeded the input-referred noise of the DVM on the 10 V range  \cite{lapuh_sampling_2018} by more than 20 dB, so the DVM contribution could be ignored.

\begin{figure}[h!] 
\centering
\includegraphics[width=8.5cm]{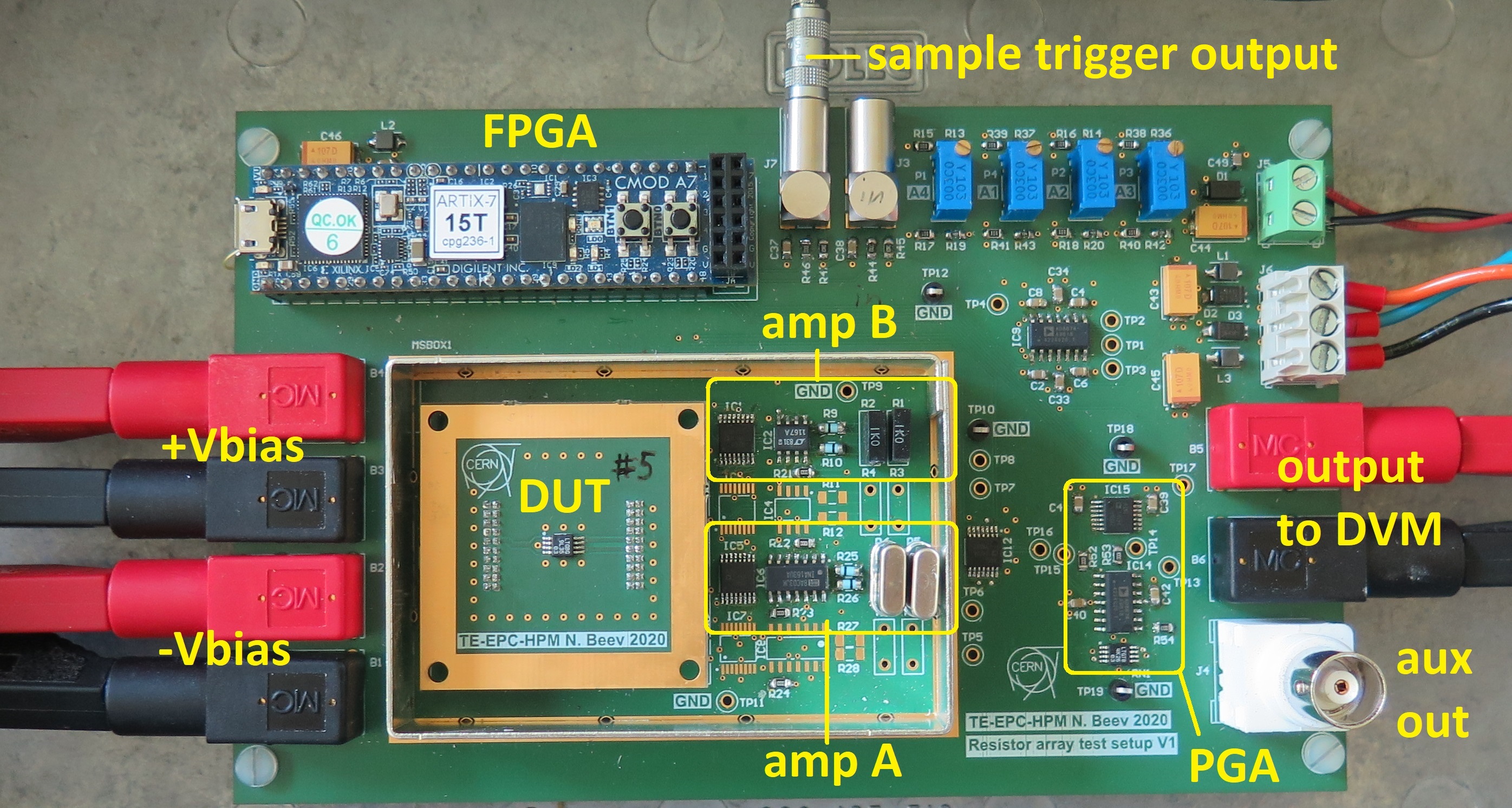}
\caption{Photograph of the mainboard with a DUT}
\label{fig:photo}
\end{figure}

The setup containing the mainboard and DUT was placed in a large die cast aluminium box. The DUT carrier board was covered with a machined aluminium lid and was enclosed in a steel shielding box together with the instrumentation amplifiers. Multiple layers of shielding were needed both for good electromagnetic interference (EMI) suppression, and to protect the DUT and the amplifiers from turbulent air flow and short-term temperature variations. Measurements were taken after the thermal settling of the DUT, which varied considerably among different devices and reached a few hours for some units. An annotated photograph of the mainboard containing a DUT is shown in Figure \ref{fig:photo}, with the shields removed for clarity.

\subsection{Signal acquisition and processing}

The most important parameters related to signal acquisition and processing are given in Table \ref{tab:signal}. 

\begin{table}
\caption{Signal acquisition and processing parameters}\label{tab:signal}
\begin{center}
 \begin{tabular}{|| l | l ||} 
 \hline
 Sampling rate & 400 Sa/s \\ 
 \hline
 Sampling aperture & 2.4 ms\\
 \hline
 Single buffer duration & 100 s \\
 \hline
 Number of FFT averages & 100 \\
 \hline
 Window function & Hanning \\ 
 \hline
\end{tabular}
\end{center}
\end{table}

The bridge signal was acquired using the correlated double sampling (CDS) method \cite{linear_an96}. This simple but powerful technique recovers the coherent signal, while suppressing parasitic near-DC components originating in the signal chain such as thermal or galvanic electromotive forces, amplifier drift, and \textit{1/f} noise. The hardware part of the implementation included reversals of the bridge bias voltage and synchronized sampling (Figure \ref{fig:blockdiagram}). The reversal rate of 400 Hz was high enough to be above the \textit{1/f} corner of voltage and current noise of the amplifiers for all tested levels of R\textsubscript{DUT}. Hence the resulting post-CDS system noise floor was always flat from the lowest resolved frequency to \textit{f\textsubscript{s}/2}.

The full buffer containing the sampled signal was read from the 3458A DVM over the GPIB interface and the software part of the CDS scheme was implemented in a quasi real-time LabView routine. The interleaved samples taken with opposite bridge bias polarities were separated into two arrays. The arrays were then subtracted element-by-element and the result normalized, effectively implementing a two-point FIR filter having the kernel (0.5; -0.5). The resulting signal had twice lower equivalent sampling rate of 200 Hz. It also contained a large DC component that was subtracted prior to the Fast Fourier Transform (FFT) processing, which was also done in LabView.

Multiple windowed FFTs were averaged to produce smooth spectra that were saved for offline processing in MATLAB. For each measurement, the amplifier noise contribution was rms-subtracted, and numerical integration yielded the rms noise in the frequency decade from 0.01 to 0.1 Hz from which NI was calculated. The noise of each element within the bridge was assumed to be independent, leading to a correction factor of $\sqrt{4}=2$. Power law fitting was carried out to extract the exponent $\alpha$ for measurements where excess noise was sufficiently well above the Johnson noise floor.

\subsection{Noise characterization}
\begin{figure}[h!] 
\includegraphics[width=8.5cm]{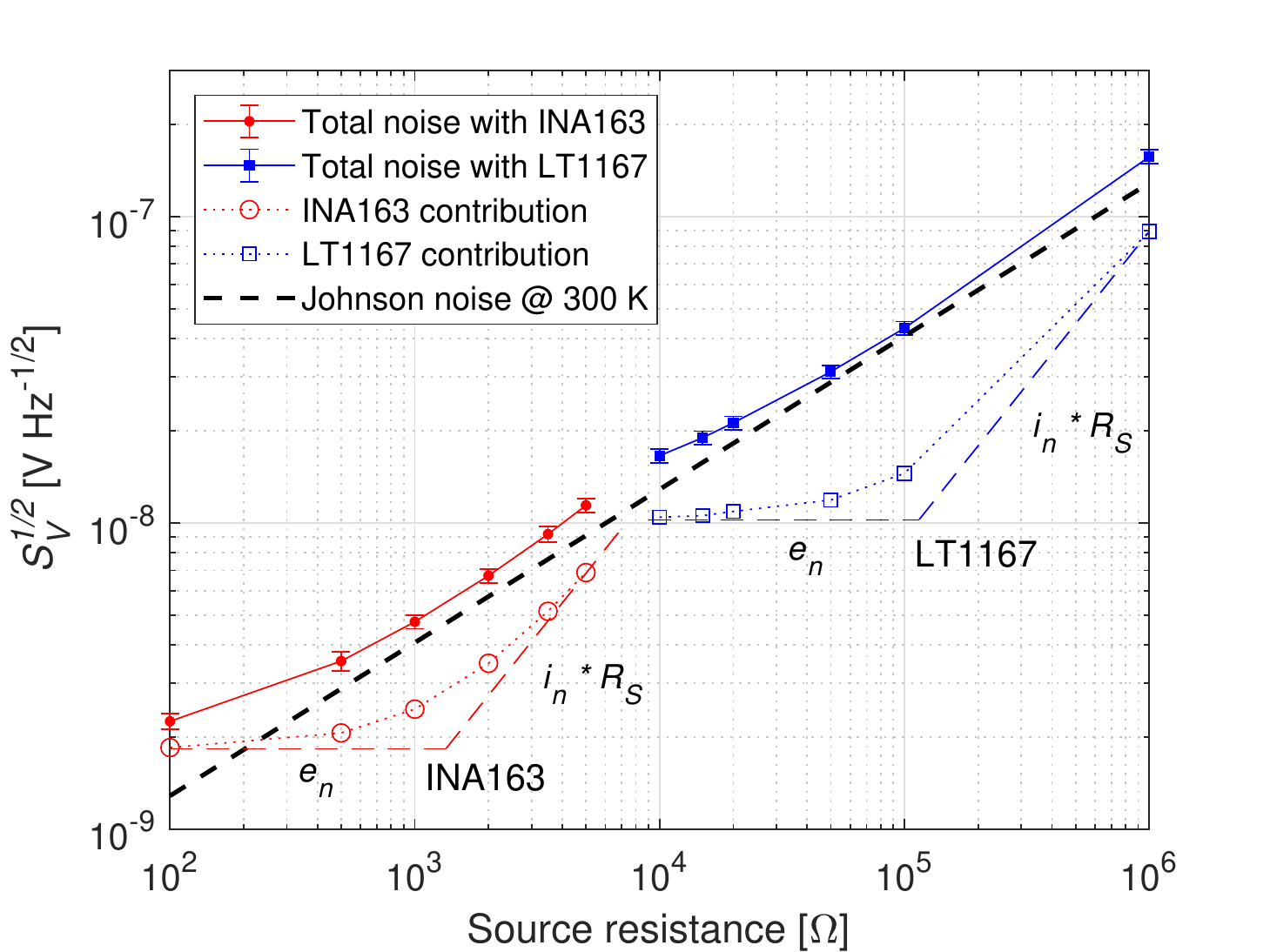}
\caption{System noise and contributions of the instrumentation amplifiers}
\label{fig:noiseamps}
\end{figure}

System noise was evaluated at each tested source resistance to estimate the contributions of the instrumentation amplifiers. The results are summarized in Figure \ref{fig:noiseamps}. Table \ref{tab:inampnoise} lists separately the measured input-referred current and voltage noise density levels of the amplifiers. The voltage noise of INA163 dominated only at the lowest source resistance of 100 $\Omega$, while in all other cases the amplifier contribution was below the Johnson noise of the DUT.

The output noise of the SIM928 voltage sources was also measured down to 0.01 Hz at each bias voltage level. Its contribution was evaluated for each measurement and was always found to be negligible due to the high common-mode rejection of the Wheatstone bridge built of well-matched precision resistors. The lowest DUT common-mode rejection ratio (CMRR) was 62 dB, while for more than 90\% of the DUTs it exceeded 80 dB.

\begin{table}[h!] 
\caption{Voltage and current noise of the instrumentation amplifiers} \label{tab:inampnoise}
\begin{center}
 \begin{tabular}{|| l | l | l ||} 
 \hline
 & \textbf{INA163} & \textbf{LT1167} \\ 
 \hline
 \textit{e\textsubscript{n}} & 1.83 nV Hz$^{-1/2}$ & 10.23 nV Hz$^{-1/2}$ \\
 \hline
 \textit{i\textsubscript{n}} & 1.37 pA Hz$^{-1/2}$ & 89.3 fA Hz$^{-1/2}$ \\
 \hline
\end{tabular}
\end{center}
\end{table}

\section{Results}

\subsection{Thin film networks}

The tested thin film networks are listed in Table \ref{tab:thinfilmresults}.

\begin{table}
\caption{List of tested thin film networks}\label{tab:thinfilmresults}
\begin{center}
 \begin{tabular}{||p{2.5cm}|p{5.5cm}||}
 \hline
 \textbf{Manufacturer} & \textbf{Resistor network family} \\ 
 \hline
 Analog Devices & \textit{LT5400}\\
 \hline
 Susumu & \textit{RM3216F} \\
 \hline
 Vishay Dale & \textit{NOMCA, AORN, DFN, TOMC, NOMC, MORN, OSOP, HTRN, ORN, TDP, MPM, VSOR} \\
 \hline
 Vishay Beyschlag & \textit{ACAS} \\ 
 \hline
 Vishay Sfernice & \textit{PRA} \\ 
 \hline
 TT Electronics & \textit{DIP-1999, DIV23} \\ 
 \hline
 Maxim & \textit{MAX549x} \\ 
 \hline
 KOA Speer & \textit{RIA} \\ 
 \hline
 Caddock & \textit{T914} \\ 
 \hline
\end{tabular}
\end{center}
\end{table}

\begin{figure}[h!] 
\includegraphics[width=8.5cm]{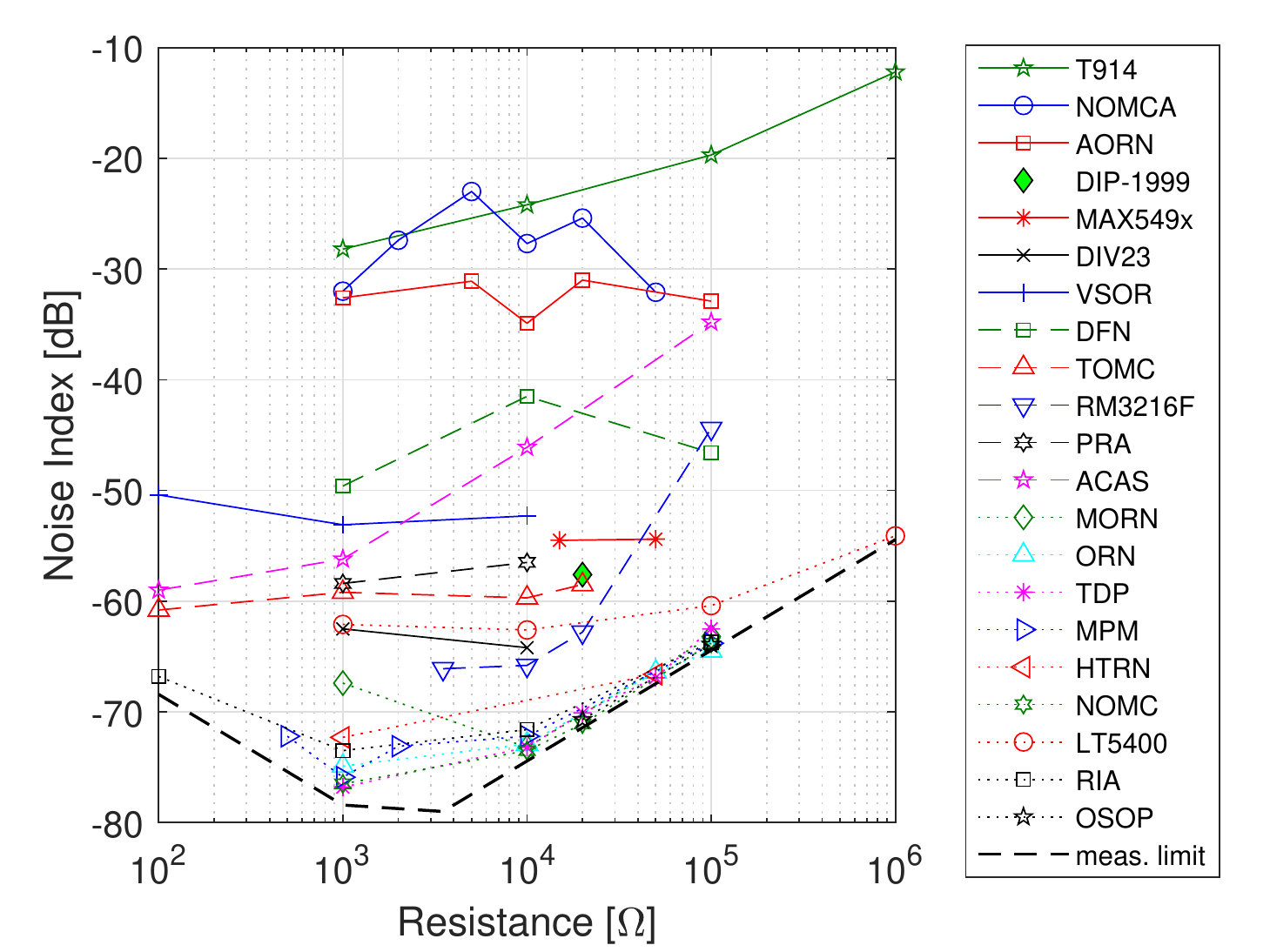}
\caption{Noise Index by family for thin film networks. Dotted lines indicate NiCr on Si substrate, dashed lines are NiCr on Alumina, and solid lines are other film types}
\label{fig:noisethinfilm}
\end{figure}

The NI results for all tested parts are shown in Figure \ref{fig:noisethinfilm}. The dashed line at the bottom represents the sensitivity limit of the measurement, defined by Johnson noise and bias voltage.

Three lots of 1 k$\Omega$ networks from the \textit{NOMCA} family with recent date codes were tested separately and found to have a spread of 8 dB (batch-to-batch) and \textless 1 dB (within a batch). One unit exhibited two-level resistance jumps with a characteristic time of seconds. Older samples of 10 k$\Omega$ \textit{TOMC} networks were also found to be noisier by about 10 dB compared to more recent ones, but the spread was not studied systematically.

The slope \textit{$\alpha$} of the \textit{1/f$^\alpha$} power spectrum density was analyzed for those DUTs where the NI exceeded the measurement floor by at least 10 dB. It was found to be in the range of 0.64 to 1.19 with a median value of 0.92. No strong correlation was seen between $\alpha$ and NI or R\textsubscript{DUT} for the tested samples.

\subsection{Metal foil networks}

The tested metal foil resistor networks from Vishay Precision Group are listed in Table \ref{tab:foilresults}. None of them exhibited any measurable excess noise down to 0.01 Hz when biased with up to 10 V per element. Hence the measurement limits shown in Figure \ref{fig:noisethinfilm} can be taken as conservative NI upper limits for these resistors.

\begin{table}[!h]
\caption{List of tested metal foil networks}\label{tab:foilresults}
\begin{center}
 \begin{tabular}{||p{3cm}|p{3cm}||}
 \hline
 \textbf{Resistor network family} & \textbf{Values} \\ 
 \hline
 \textit{SMN} & 10 k$\Omega$\\
 \hline
 \textit{SMNZ} & 1 k$\Omega$, 5 k$\Omega$, 10 k$\Omega$ \\
 \hline
 \textit{VHD200} & 20 k$\Omega$ \\
 \hline
 \textit{PRND-1446} \cite{prnd1446} & 3.5 k$\Omega$, 20 k$\Omega$   \\ 
 \hline
\end{tabular}
\end{center}
\end{table}

\begin{figure}[h!] 
\includegraphics[width=8.5cm]{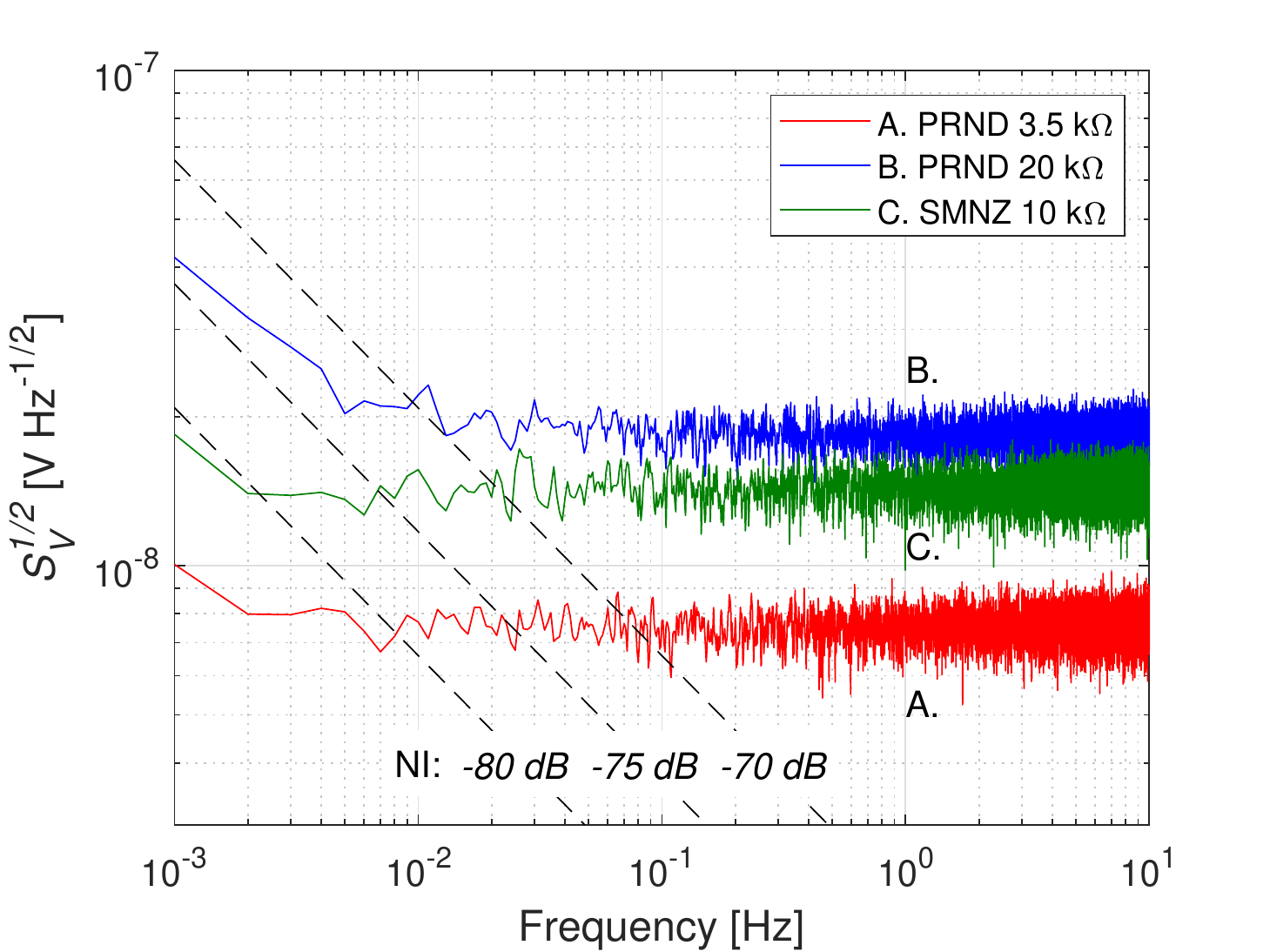}
\caption{Noise spectral density for some metal foil resistors biased with 10 V per element}
\label{fig:noisefoil}
\end{figure}

Additional measurements were taken to study some metal foil resistors and to explore the practical limitations of the setup. Instead of the SIM928 voltage sources, two low-noise 10 V standards based on LTZ1000 \cite{fernqvist_design_2003} were used for biasing. The temperature of the DUT was stabilized to mK levels using a PT1000 sensor and a thermoelectric cooler. The buffer length was increased 10-fold to allow for FFT start frequency and resolution of 1 mHz. Figure \ref{fig:noisefoil} shows the voltage noise spectral density of three different tested parts, together with NI contours down to -80 dB.

\subsection{Summary}

\begin{figure}[h!] 
\includegraphics[width=8.5cm]{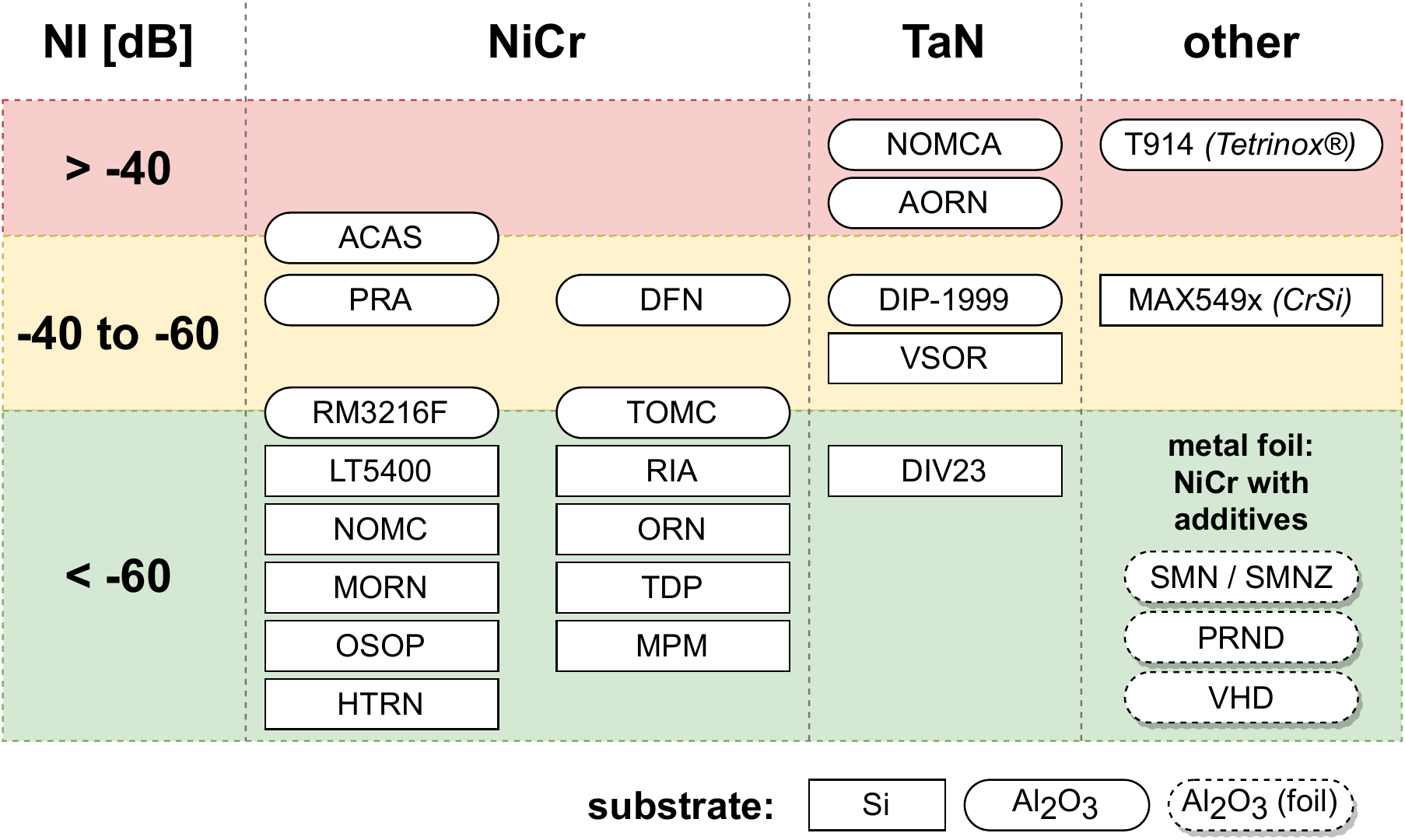}
\caption{Summary for NI of all tested resistor network families}
\label{fig:summary}
\end{figure}

The results for all tested resistor network families are summarized in Figure \ref{fig:summary}.

\section{Discussion}

The measurements presented in this work show that many types of precision resistor networks have extremely low levels of excess noise well below 1 nV\textsubscript{rms}/V\textsubscript{DC}/decade (NI \textless -60 dB). Both metal foil and thin film NiCr networks deposited on silicon substrates can be used in circuits that aim for very low-noise performance. Some components with Alumina substrates also show low levels of excess noise that would suit most practical applications.

Among the tested devices having intermediate and high excess noise levels, only a few types exhibited a strong dependence of NI on resistance (see Figure \ref{fig:noisethinfilm}). This finding was unexpected as it is contradictory to the usual assumption that for a given resistor type NI increases with resistance as the volume of the resistive element decreases \cite{zandman} \cite{motchenbacher} \cite{seifert}. It could be indicative of dominant noise mechanisms originating outside the film, e.g. in the contacts \cite{rolke_nichrome_1981} \cite{fisher_termination_1971} or in points of high local current density such as trim links \cite{vandamme_2002}.

Numerous studies have linked the electrical properties of thin film resistors to physical characteristics of the substrate. R\"{o}lke demonstrated the strong relation between substrate surface roughness and resistance spread in NiCr resistors \cite{rolke_nichrome_1981}. Lai \textit{et al.} studied NiCr thin films deposited on glass, silicon and copper substrates, and found that surface roughness had a considerable impact on resistivity and very little effect on TC \cite{lai_comparison_2013}. Motchenbacher and Fitchen argued that high-resistance metal films are more strongly affected by surface variations in the context of excess noise \cite{motchenbacher}.

The results of the hereby presented study suggest a strong link between excess noise and substrate type for NiCr thin film resistor networks. Semiconductor-grade silicon substrates have very smooth surfaces with roughness on the nm scale \cite{lai_comparison_2013} and all tested DUTs having such substrates exhibited low noise (Figure \ref{fig:summary}). On the contrary, Alumina substrates have much higher surface roughness that typically exceeds the thickness of the deposited thin films. If the manufacturing processes are assumed to be well controlled and repeatable, wafer-to-wafer roughness variation could be a possible explanation for the NI spread seen between different batches or between different values within the same family (see the plots for \textit{NOMCA}, \textit{AORN} and \textit{DFN} in Figure \ref{fig:noisethinfilm}).
The limited number of available parts of TaN and other technologies prevents us from drawing more general conclusions.

The noise spectra of the tested parts were found to be of regular \textit{1/f$^\alpha$} type, with the exception of one unit that exhibited two-level burst noise. The exponent \textit{$\alpha$} could not be related to other parameters. Its distribution had a maximum at values slightly below 1, which confirms earlier findings on large ensembles of thin film resistors from multiple manufacturers \cite{sikula_1/f_1996} \cite{hruska_thin_1997}. The suitability of NI as a quantitative indicator should be carefully considered for parts where \textit{$\alpha$} deviates strongly from unity and where the noise power is no longer equal per log unit of frequency bandwidth \cite{vandamme_2002}.

In metal foil resistors the ceramic substrate plays a crucial role for TC compensation, but clearly has no impact on the microstructural properties of the resistive element, the contacts or the trim links, because 1) the foil is formed separately before it is bonded to the substrate; 2) the attachment is done via a thick layer of adhesive (a few $\mu$m), and 3) the foil is also very thick ($\mu$m-scale) compared to thin films \cite{zandman}. Therefore, for this type of resistors the extremely low NI can be explained by the bulk conductivity mechanism of the foil.

Past research has explored the possibility of using excess noise as an indicator of resistor quality, reliability, or long-term irreversible drift \cite{zandman} \cite{hruska_thin_1997}. Anomalously high levels have been associated with defects in thin films \cite{vossen_screening_1973} or damage due to electromigration \cite{cottle_microstructural_1990} \cite{koch_relationship_1993}, chemical or electrochemical corrosion \cite{rolke_nichrome_1981}. The methods and results presented in this work could serve as a reference and starting point for further investigations on resistor properties via low-frequency noise measurements.

\section*{Acknowledgment}

The author is grateful to Susumu, KOA Speer and Caddock for providing free samples and support. The technical help of Vishay Dale, Maxim and  Analog Devices is also acknowledged. Special thanks to John R. Pickering of Metron Designs for the useful discussions. The manuscript was prepared with the essential help of Miguel Bastos and Michele Martino of CERN.

This research has been supported by the HL-LHC project.

\ifCLASSOPTIONcaptionsoff
  \newpage
\fi

\bibliographystyle{unsrt}
\bibliography{references} % Entries are in the "refs.bib" file

\appendices
\section{Noise Index Nomogram}

The nomogram given in this Appendix can be used to quickly calculate various quantities related to excess noise. For a given NI, voltage bias and resistance, it can be used to determine the \textit{1/f} corner frequency \textit{f\textsubscript{c}}. It is also possible to use it inversely, in case the designer wants to find the NI required for a known resistance, voltage bias, and \textit{f\textsubscript{c}}. The example isopleth lines are drawn for a part having R=1 k$\Omega$ and NI=-50 dB biased with V\textsubscript{DC}=10 V. The Johnson noise levels given at the rightmost scale are calculated for T=300 K.

\begin{wrapfigure}{l}{18cm} 
\begin{framed}
\includegraphics[width=17cm]{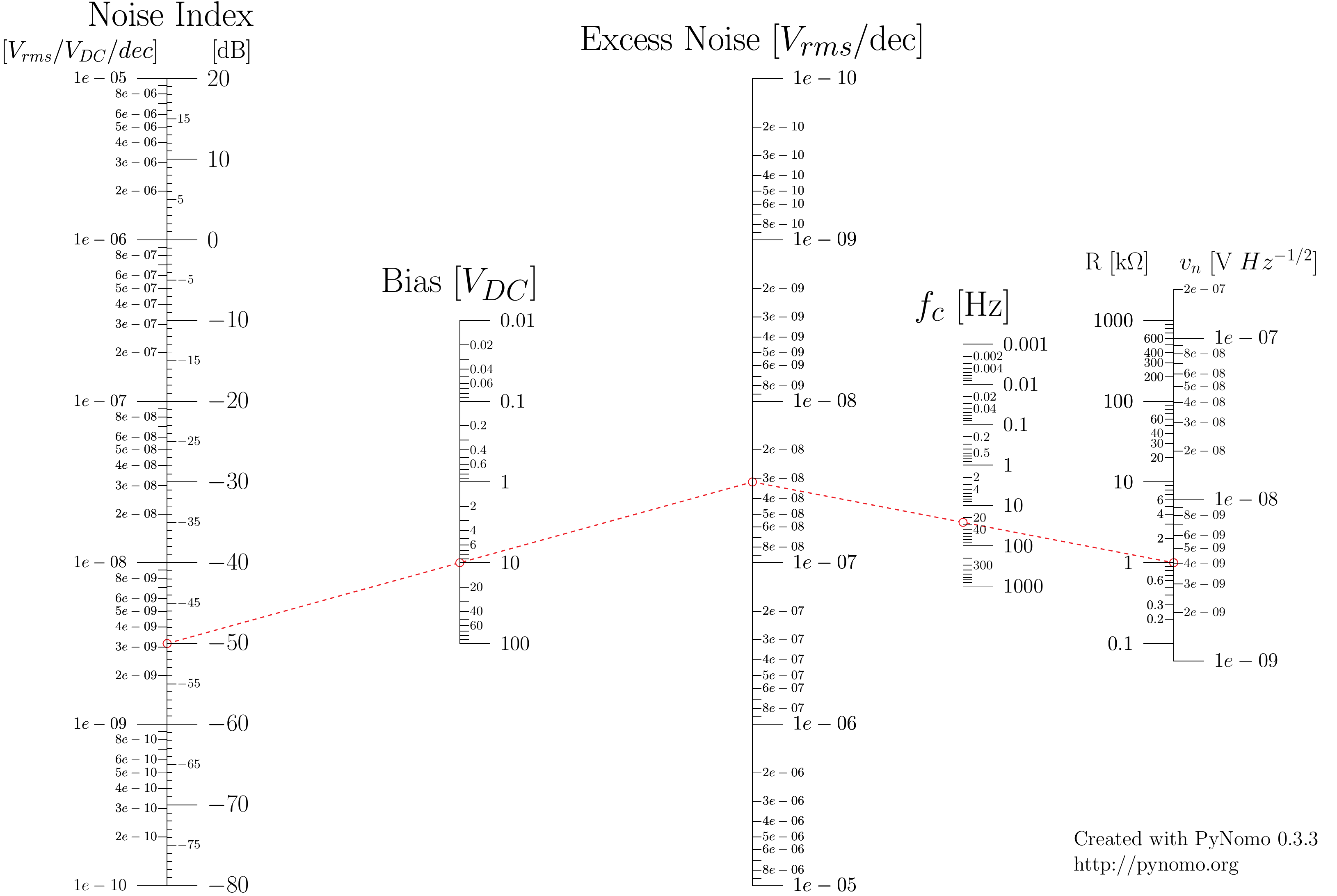}
\label{fig:nomogram}
\end{framed}
\end{wrapfigure}

\end{document}